\documentclass{article}

\usepackage{arxiv}

\usepackage[utf8]{inputenc} 
\usepackage[T1]{fontenc}    

\usepackage{siunitx}
\usepackage{algorithm}
\usepackage{algpseudocode}
\usepackage{bbm}
\usepackage{booktabs}
\usepackage{multirow}
\usepackage{xcolor}
\usepackage{mathtools}
\usepackage{amsmath}
\usepackage{enumitem}

\usepackage[hidelinks]{hyperref}
\usepackage{url}
\hypersetup{breaklinks=true}
\usepackage{cite}
\usepackage{stix}
\usepackage{svg}

\newcommand{\bd}[1]{\boldsymbol{#1}}
\usepackage{placeins}

\usepackage{booktabs}       
\usepackage{amsfonts}       
\usepackage{nicefrac}       
\usepackage{microtype}      
\usepackage{lipsum}		
\usepackage{graphicx}
\usepackage{natbib}
\usepackage{doi}

\title{ISLAND: Interpolating Land Surface Temperature using land cover}

\author{
    Yuhao Liu \\
    Rice University \\
    \texttt{yuhao.liu@rice.edu}
    \And Pranavesh Panakkal \\
    Rice University \\
    \texttt{pranavesh@rice.edu}
    \And Sylvia Dee \\
    Rice University \\
    \texttt{sylvia.dee@rice.edu}
    \And Guha Balakrishnan \\
    Rice University \\
    \texttt{guha@rice.edu}
    \And Jamie Padgett \\
    Rice University \\
    \texttt{jamie.padgett@rice.edu}
    \And Ashok Veeraraghavan \thanks{Corresponding author}\\
    Rice University \\
    \texttt{vashok@rice.edu}
}

\renewcommand{\shorttitle}{Liu et al.}

\hypersetup{
pdftitle={ISLAND: Interpolating Land Surface Temperature using land coverote Sensing},
pdfauthor={Yuhao Liu, Pranavesh Panakkal, Sylvia Dee, Guha Balakrishnan, Jamie Padgett, Ashok Veeraraghavan},
pdfkeywords={cloud removal, land surface temperature, thermal imaging, Landsat, land cover},
}

\begin{document}
\maketitle

\begin{abstract}
Cloud occlusion is a common problem in the field of remote sensing, particularly for retrieving Land Surface Temperature (LST). Remote sensing thermal instruments onboard operational satellites are supposed to enable frequent and high-resolution observations over land; unfortunately, clouds adversely affect thermal signals by blocking outgoing longwave radiation emission from the Earth's surface, interfering with the retrieved ground emission temperature. Such cloud contamination severely reduces the set of serviceable LST images for downstream applications, making it impractical to perform intricate time-series analysis of LST. In this paper, we introduce a novel method to remove cloud occlusions from Landsat 8 LST images. We call our method ISLAND, an acronym for \underline{I}nterpolating Land \underline{S}urface Temperature using \underline{land} cover. Our approach uses LST images from Landsat 8 (at \SI{30}{\meter} resolution with 16-day revisit cycles) and the NLCD land cover dataset. Inspired by Tobler's first law of Geography, ISLAND predicts occluded LST through a set of spatio-temporal filters that perform distance-weighted spatio-temporal interpolation. A critical feature of ISLAND is that the filters are land cover-class aware, making it particularly advantageous in complex urban settings with heterogeneous land cover types and distributions. Through qualitative and quantitative analysis, we show that ISLAND achieves robust reconstruction performance across a variety of cloud occlusion and surface land cover conditions, and with a high spatio-temporal resolution. We provide a public dataset of 20 U.S. cities with pre-computed ISLAND LST outputs. Using several case studies, we demonstrate that ISLAND opens the door to a multitude of high-impact urban and environmental applications across the continental United States.
\end{abstract}



\section{Introduction}
\label{sec:1_intro}
Land surface temperature (LST) is a fundamental aspect of Earth's climate system, it attracts extensive studies across diverse disciplines. The wide array of fields include climate change \citep{horton2016review,IPCC2021extremes}, urban planning \citep{sobrino_evaluation_2013,huang_investigating_2019,osborne2019quantifying}, vegetation and land cover changes~\citep{gomez-martinez_multi-temporal_2021,chun2018impact}, and human health \citep{Orimoloye_Mazinyo_Nel_Kalumba_2018,dee2022increasing,dugord2014land}. LST changes rapidly both in space and time due to the strong heterogeneity of land surface characteristics and the short timescales of weather. As a consequence, accurate characterization of LST requires dense spatial and temporal sampling \citep{li_satellite-derived_2013}.

\begin{figure}
    \centering
    \includegraphics[width=0.9\linewidth]{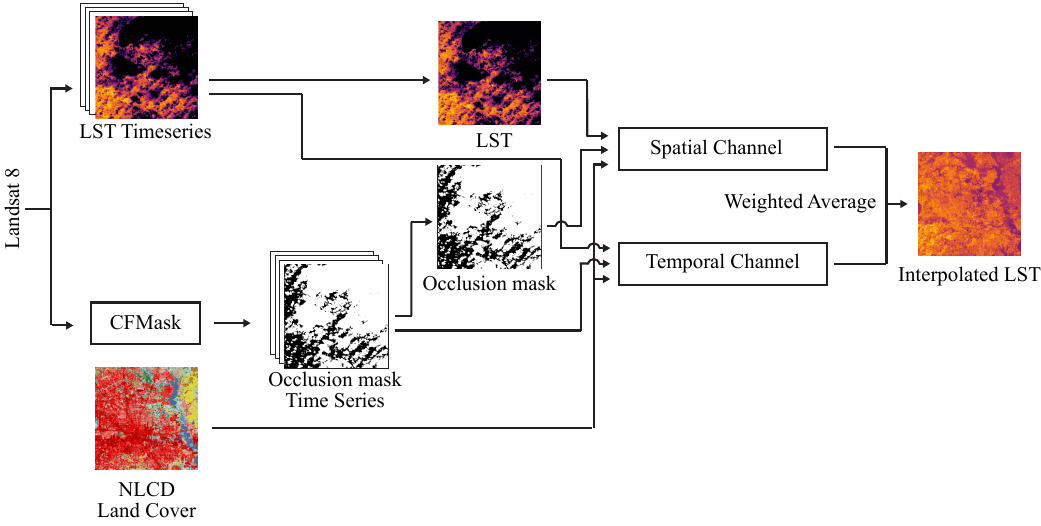}
    \caption{\textbf{Schematic overview.} A block diagram of our proposed method ISLAND. We use LST images from Landsat 8 and land cover labels from NLCD as inputs. We identify cloud-contaminated pixels via CFMask and build an occlusion mask for each LST image. ISLAND performs class-aware interpolation using two channels: spatial channel and temporal channel. The interpolated LST output is a weighted average between the two channels.}
    \label{fig:overview_flowchart}
\end{figure}

Growing recognition of the importance of accurate LST observations has driven
rapid advances in remote sensing~\citep{li_satellite-derived_2013}.
Satellite-based thermal infrared (TIR) data provides LST measurements with high spatial and temporal resolution at a global scale.
As an example, the National Aeronautics and Space Administration (NASA) Landsat
8 satellite was launched in 2013, and its platform has provided detailed LST
data over the last decade~\citep{landsat8,landsat9}.
The Landsat 8 and 9 satellites each include a Thermal Infrared Sensor (TIRS) onboard, which collects LST data at a spatial resolution of \SI{30}{\meter} with a revisit cycle of 16 days.
The high spatial resolution of TIRS has enabled numerous studies of LST over
heterogeneous regions like urban
centers~\citep{huang_investigating_2019,sobrino_evaluation_2013,streutker_satellite-measured_2003}
and wetlands~\citep{eisavi_spatial_2016,demarquet_long_2023}.

Despite these advancements, cloud occlusion persists as a substantial obstacle to achieving reliable spaceborne retrievals of LST.
Clouds adversely affect LST readings by blocking thermal radiation emitted from Earth's surface.
This results in cloud-contaminated pixels exhibiting considerably lower values
compared to their true values, rendering the affected images
unusable~\citep{jin_cloud_2019}.
Unfortunately, cloud contamination is a frequent phenomenon in Landsat images,
with studies indicating that an average of 35\% of Landsat images globally
contain missing data due to cloud
contamination~\citep{roy_multi-temporal_2008}.
Moreover, most cities experience enhanced daytime cloud cover compared to their
surrounding rural regions~\citep{vo2023urban}, making cloud-free LST images
even harder to obtain for cities.
Furthermore, regions at higher latitudes, such as Pacific Northwest cities like
Portland and Seattle, are prone to rainier weather conditions,
leading to more frequent cloud contaminations~\citep{noaa2020comparative}.
Under these circumstances, obtaining a single cloud-free image may be infeasible for weeks or even months, severely limiting the utility of Landsat LST data.
Collectively, these barriers have limited practical high spatial and temporal sampling of LST. Thus, the temporal dynamics of LST over complex, heterogeneous terrains remain under-observed, under-studied, and under-constrained, especially at scale.

In this paper, we present a novel method to mitigate the effects of cloud contamination in satellite LST images.
Our method incorporates Tobler's First Law of Geography (TFL), which states
that ``everything is related to everything else, but near things are more
related than distant things''~\citep{tobler_1970}.
In addition to the distance-decay effect from TFL, we also incorporate multi-temporal information and, crucially, integrate land cover data into our model.
Land cover data contains information about physical land types, such as forests, open water, and urban/developed areas.
While existing studies have demonstrated the strong relationship between land
cover types and
LST~\citep{Chaudhuri_Mishra_2016,zhao_assessing_2020,imran_impact_2021}, our
model represents the first attempt to utilize land cover to infer occluded LST
pixel values in satellite images.
We call our model ISLAND, an acronym for \underline{I}nterpolating Land \underline{S}urface Temperature using \underline{land} cover. 
ISLAND performs interpolation using a set of spatio-temporal filters to capture surrounding pixel values and historical patterns. 
Notably, these filters are designed to be sensitive to land cover classes, enabling class-specific pattern capture and higher reconstruction accuracy. 

We demonstrate that ISLAND provides robust LST reconstruction performance across various occlusion and land cover conditions.
Our results indicate that ISLAND greatly improves the practical temporal resolution of Landsat LST images by robustly estimating cloud-contaminated LST pixels.
Simulation evaluation shows that the RMSE in our reconstructed Landsat 8 LST images is around $1.65-$\SI{2.62}{\kelvin}. In situ evaluation shows that our RMSE is approximately $3.90-$\SI{4.75}{\kelvin}. 
We present three illustrative examples underscoring the utility of ISLAND, namely, (1) urban heat island effects, (2) derivation of surface temperature trends, and (3) social vulnerability and urban heat stress.

ISLAND leverages publicly available data from Landsat 8~\citep{landsat8} and
National Land Cover Database (NLCD)~\citep{yang_new_2018}. This approach ensures the accessibility and
transferability of ISLAND, as it is open-source and can be easily deployed in
any region within the continental United States (CONUS) as per user
requirements.
All source code\footnote{https://github.com/Way-Yuhao/ISLAND} is available to the public, accompanied by a dataset\footnote{https://doi.org/10.17603/ds2-3rf5-sd58} comprising 20 urban regions with pre-computed ISLAND LST outputs.
We envision that ISLAND will provide tremendous operational value for a variety of applications in Earth, geospatial, and social sciences, and ultimately pave the way for a new generation of LST studies using remote sensing.

\section{Related Work}
\label{sec2:related}
The refinement of cloud-removing algorithms is an open area of research in the field of remote sensing.
Several existing studies develop and discuss cloud removal algorithms for remotely acquired LST. 
Unfortunately, these algorithms generally target lower spatial resolutions and
homogeneous land cover types~\citep{wu_spatially_2021}.
For example,~\citet{rs11030336} proposed a method to reconstruct Moderate
Resolution Imaging Spectroradiometer (MODIS) LST over cloudy pixels using land
energy balance theory and similar pixels at a spatial resolution of
\SI{1}{\kilo\meter}.
\citet{6891149} proposed a spatiotemporal technique to reconstruct MODIS LST
products using regression and multispectral ancillary data to classify pixels,
which improves reconstruction accuracy.
Unfortunately, MODIS LST at \SI{1}{\kilo\meter} is too coarse to resolve urban infrastructures, such as buildings and roads.

There are existing studies that reconstruct cloudy-sky LST at the spatial resolution of Landsat 8 pixels (\SI{30}{\metre}).
\citet{wang2019recovering} proposed a method to reconstruct cloud-sky Landsat 8
LST by considering solar-cloud-satellite geometry.
However, this method requires a temporally adjacent clear-sky image as a reference, which is hard to obtain, limiting its robustness.
Furthermore, the accuracy of their reconstructed Landsat LST was not validated against in situ LST measurements.
\citet{ZHU2022113261} reconstruct Landsat 8 LST data using an annual
temperature cycle (ATC) model and adjacent spatial information from similar
pixels. 
\citet{ZHU2022113261} validated their results against in situ LST measurements
at six Surface Radiation Budget Network (SURFRAD) sites~\citep{surfrad_2000}.
However, there are several limitations to their study.
First, their reconstruction method may fail when the cloud cover percentage exceeds 70\%, limiting the scope of their application.
Additionally, the ATC model assumes a constant mean annual surface temperature per region, suggesting that their model likely underestimates LST under global warming and the increased prevalence of the urban heat island effect.
Lastly, the primary limitation for~\citet{ZHU2022113261} and all existing
studies mentioned in this section is that these studies are mainly designed and
validated on relatively homogeneous regions of land cover, such as cropland,
shrubland, and grassland; their results were not demonstrated over urban
regions.
This study addresses this gap and provides a scalable method to generate high-resolution LST, even for highly urbanized regions with complex temperature feedback and heterogeneous urban land surface types.
We achieve this by explicitly employing NLCD land cover labels and using the satellite LST product with the highest resolution (Landsat 8) to maximize its usability in cities.

 Our model shares some of the high-level design principles with existing models, such as the use of multitemporal remote sensing data.
To the best of our knowledge, ISLAND is the first algorithm that incorporates land cover labels as ancillary data for cloud removal in LST images.
We summarize our \textit{key contributions} as follows: 
(1) We use NLCD land cover labels to accurately reconstruct Landsat 8 LST under cloudy-sky conditions.
(2) Using spatial adjacency and multi-temporal filters, ISLAND effectively removes cloud contaminations from Landsat 8 LST images even under severe occlusion, thereby dramatically improving the temporal resolution of Landsat 8 LST products.
(3) Simulation evaluation shows that the RMSE in our reconstructed Landsat 8 LST images is around $1.65-$\SI{2.62}{\kelvin}. In situ evaluation shows that our RMSE is approximately $3.90-$\SI{4.75}{\kelvin}. 
(4) We release a public dataset of cloud-free, reconstructed LST maps for 20 US cities from 2019--2023. 

\section{Data}
\label{sec:3_data}

In this section, we explain the implementation of the data compilation steps of our model. See Fig.~\ref{fig:overview_flowchart} for a visual overview.

\subsection{Landsat 8 Data}

We collect Landsat 8 Collection 2 Level-2 LST products from Google Earth
Engine~\citep{gorelick2017google}.
Landsat 8 provides LST products at \SI{30}{\meter} resolution and a revisit cycle of 16 days.
Landsat 8 collects top-of-atmosphere (TOA) spectral radiance values via its band 10 Thermal Infrared Sensor (TIRS).
LST is derived from TOA spectral radiance using the radiative transfer equation
(RTE)-based single-channel algorithm~\citep{malakar2018operational}.
ASTER GED~\citep{hulley_aster_2015} was used to correct the effect of surface
emissivity.
Atmospheric effects were compensated by the Goddard Earth Observing System, Version 5 (GEOS-5) reanalysis data using the radiative transfer model MODTRAN 5.2. 
The RMSE of Landsat LST products is approximately
\SI{2.2}{\kelvin}~\citep{malakar2018operational}.
Please refer to~\citet{malakar2018operational} for a more detailed description
of the Landsat LST retrieval algorithm.
We denote the input Landsat LST image as $\Tilde{T} \in \mathbb{R}^{H \times W}$, where $H$ and $W$ denote the height and width of the LST image, respectively.

The Landsat LST product includes cloud mask information
(CFMask)~\citep{zhu2012object} on a per-pixel basis.
CFMask provides pixel quality attributes for each Landsat 8 LST image, indicating the presence of cloud contamination. 
We build a pixel-wise binary occlusion mask $O \in \{0, 1\}^{H \times W}$ for each Landsat LST image, where the pixel value is set to True if CFMask indicates that there is cloud, cloud shadow, or cirrus.
From the occlusion mask, we calculate the \emph{occlusion factor} $\theta \in [0, 1]$ that measures the fraction of pixels occluded. 
Formally, we have $\theta = \frac{\sum_{\bd{p}} O_{\bd{p}}}{H \cdot W}$, where $O_p$ denotes a pixel of $O$.
$\theta$ is an important metric that measures the severity of cloud contamination.
A higher $\theta$ indicates more severe occlusion.

\subsection{NLCD Land Cover data}

For land cover data, we use the National Land Cover Database
(NLCD)~\citep{yang_new_2018} from the U.S. Geological Survey (USGS).
NLCD provides spatially referenced descriptive data on the characteristics of the land surface using a set of thematic classes (e.g.,~urban, forest, and agriculture).
NLCD is available for the CONUS region at \SI{30}{\meter} resolution with a release cycle of once per 3~years.
We denote the NLCD land cover image as $L \in \mathbb{Z}^{H \times W}$.

\subsection{In situ data}

We collect in situ LST data at four Surface Radiation Budget Network (SURFRAD)
sites~\citep{augustine2000surfrad}.
We use SURFRAD in situ data to validate our reconstruction results under cloudy conditions.
The RMSE of SURFRAD in situ LST measurements is approximately
0.5--\SI{0.8}{\kelvin}~\citep{wang2009evaluation}.
SURFRAD measures downwelling and upwelling longwave flux every minute.
The Stefan--Boltzmann law states
\begin{equation}
 \label{eq:boltzmann_law}
F^\uparrow = (1 - \epsilon)F^\downarrow + \epsilon_b \sigma T_s^4
\end{equation}
where $F^\uparrow$ and $F^\downarrow$ are measured upwelling and downwelling longwave flux, respectively, $\epsilon_b$ is the broadband longwave surface emissivity, $\sigma$ is the Stefan--Boltzmann constant, and $T_s$ is the surface skin temperature (equivalent to LST).

The in situ LST is obtained by inverting the upwelling component of Eq.~(\ref{eq:boltzmann_law})

\begin{equation}
 \label{eq:in_situ_lst}
T_s = \left\{ \frac{1}{\varepsilon_b \sigma} \left[ F^\uparrow - (1 - \epsilon)F^\downarrow \right] \right\}^{0.25}
\end{equation}

Following~\citet{malakar2018operational}, we estimate the broadband emissivity
($\epsilon_b$) via a spectral-to-broadband regression
relationship~\citep{ogawa2008estimating} using emissivity values from ASTER
GEDv3 product~\citep{hulley_aster_2015}
\begin{equation}
 \label{eq:broadband_emis}
\epsilon_b = 0.128 + 0.014\epsilon_a^{10} + 0.145\epsilon_a^{11} + 0.241\epsilon_a^{12} + 0.467\epsilon_a^{13} + 0.004\epsilon_a^{14}
\end{equation}
where $\epsilon_a^{j}$, $j = 10 - 14$ are narrowband ASTER emissivities centered on 8.3, 8.6, 9.1, 10.6, and \SI{11.3}{\micro\metre}, respectively. 

\section{Methods}
\label{sec:4_methods}
After acquiring the required inputs from Sec.~\ref{sec:3_data}, we perform interpolation to predict cloud-occluded pixels. 
Our interpolator uses two complementary mechanisms: a spatial channel and a temporal channel.

\begin{figure}
    \centering
    \includegraphics[width=0.95\linewidth]{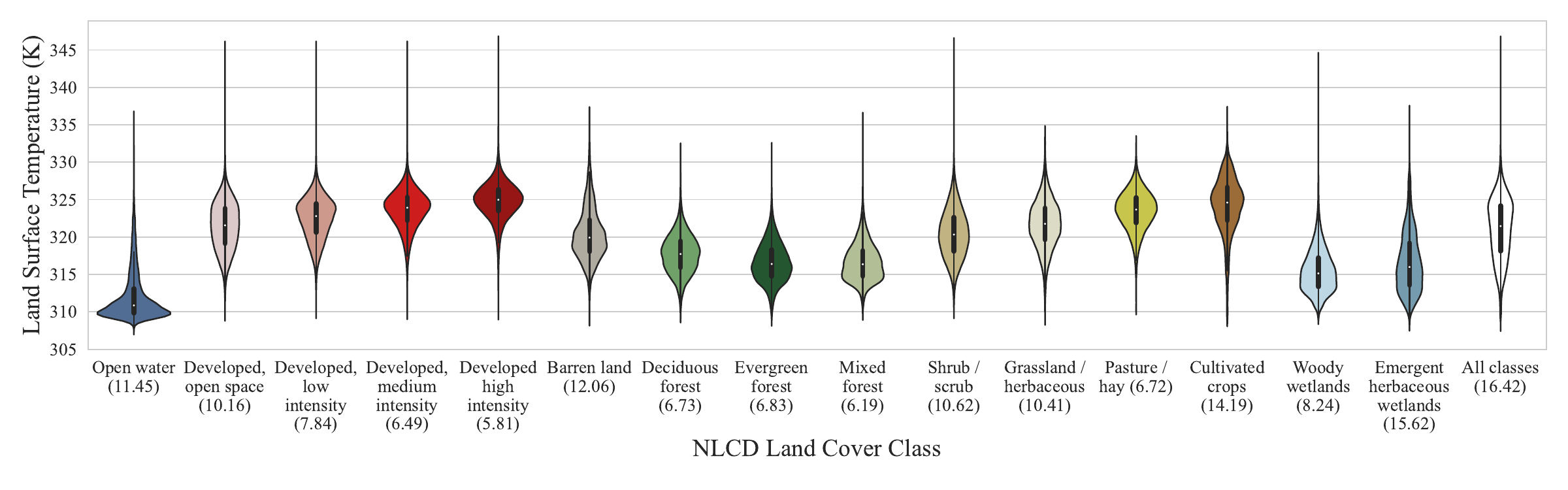}
    \caption{\textbf{Distribution of LST per land cover class.} The violin diagram shows the distribution of LST of San Antonio on August 16th, 2019. 
    The distribution of LST is dependent on land cover, and the variances (in parenthesis) conditioned on individual classes are typically lower than the variance of the parent distribution. 
    When explicitly filtering by land cover class and also by proximity, our spatial channel leads to accurate predictions of LST.}\label{fig:spatial_motivation}
\end{figure}

\subsection{Spatial Channel}
\label{subsec:spatial}
In our approach, the prediction of occluded pixel values is informed by leveraging the information embedded in their surroundings. 
We term this approach the \textit{spatial channel}.
The intuition behind the spatial channel is that nearby objects that are in the same land cover class are likely to exhibit similar thermal properties. 
This approach filters the data based on land cover class, since the distribution of LST is dependent on land cover, as depicted in Fig.~\ref{fig:spatial_motivation}.
The computation of the spatial channel bears a close resemblance to bilateral
filtering~\citep{paris2009bilateral}.
While traditional bilateral filtering considers variations in pixel intensities of the input image with the aim of preserving sharp edges, the spatial channel uses a different approach, where we filter the input image using the pixel attributes informed by auxiliary images (land cover class $L$ and occlusion $O$).

Consider our goal to be predicting $\tilde{T}$ at an occluded pixel location $\boldsymbol{p}$.
Under a low occlusion factor $\theta$, we estimate $\tilde{T}_{\bd{p}}$ using a weighted average estimate from other pixels.
Conceptually, the weights are determined based on three factors: proximity, land cover class label, and cloud occlusions.

We use a 2D Gaussian filter to model the distance-decay effect. 
Consider a pixel $\boldsymbol{q}$ within a $f \times f$ neighborhood $N$ centered at $\boldsymbol{p}$.
Let $x = \lVert \bd{p} - \bd{q} \rVert$ be the Euclidean distance between $\bd{p}$ and $\bd{q}$.
We compute the proximity weight $G_{\sigma}(x)$ using a Gaussian kernel with standard deviation $\sigma = f/2$:
\begin{equation}
G_{\sigma}(x) = \frac{1}{2 \pi \sigma^2} \text{exp}\biggl(-\frac{x^2}{2\sigma^2}\biggr) \label{eq:gaussian}
\end{equation}
$G_{\sigma}(x)$ decreases the influence of distance pixels while prioritizing the influence of nearby pixels.

We further modify the Gaussian filtering so that we only consider $\boldsymbol{q}$ that is cloud-free and has the same land cover class as $\boldsymbol{p}$.
Let $\bar{O}$ be the inverse of the occlusion mask $O$, so that $\bar{O}_{\boldsymbol{q}} = 1$ if there is no cloud contamination at $\boldsymbol{q}$, and $\bar{O}_{\boldsymbol{q}} = 0$ otherwise.
To constrain land cover class, let function $f_L$ evaluate the land cover class of two pixels $\bd{p}$ and $\bd{q}$ such that $f_L(\bd{p}, \bd{q}) = 1$ if the corresponding land cover classes of $\bd{p}$ and $\bd{q}$ are the same (i.e.,~$L_{\bd{p}} = L_{\bd{q}}$), and $f_L(\bd{p}, \bd{q}) = 0$ otherwise.
Formally, we write our weighted average local filter as follows:
\begin{equation}
\tilde{T}_{\boldsymbol{p}} = \frac{1}{\alpha} \sum_{\boldsymbol{q} \in N} G_{\sigma}(\lVert \boldsymbol{p} - \boldsymbol{q} \rVert) ~\bar{O}_{\boldsymbol{q}}~f_L(\boldsymbol{p}, \boldsymbol{q})~\tilde{T}_{\boldsymbol{q}} \label{eq:spatial_local_filter}
\end{equation}
 where $\alpha$ is a normalization parameter that ensures weights sum to $1$ within each neighborhood (i.e., $\forall \bd{q} \in N$):
\begin{equation}
\alpha = \sum_{\bd{q} \in N} G_{\sigma}(\lVert \boldsymbol{p} - \boldsymbol{q} \rVert) ~\bar{O}_{\boldsymbol{q}}~f_L(\boldsymbol{p}, \boldsymbol{q}) \label{eq:spatial_gaussian_norm}
\end{equation}

Local filtering (defined in Eq.~(\ref{eq:spatial_local_filter})) works well when the occlusion factor is low. 
For a high $\theta$, there are fewer neighboring pixels available, and local filtering leads to noisy or even invalid estimations.
Instead of local filtering, we resort to global averaging when encountering high $\theta$.
Here we estimate the pixel value of $\bd{p}$ with the average temperature of all non-occluded pixels that have the same land cover class, that is
\begin{equation}
\mu_c = \frac{1}{T} \sum_{\bd{q} \in \tilde{T} \setminus \{\bd{p}\}}\bar{O}_{\boldsymbol{q}}~f_L(\boldsymbol{p}, \boldsymbol{q})~\tilde{T}_{\boldsymbol{q}}  \label{eq:global_class_mean}
\end{equation}
 where $c$ is the land cover class of the occluded pixel $\bd{p}$, and $T$ is the total number of non-occluded pixels with the same land cover class as $\bd{p}$, excluding $\bd{p}$ itself.
Contrary to local filtering (Eq.~(\ref{eq:spatial_local_filter})), here the averaging is performed across the entire image, rather than across some local $f \times f$ window.
As a result, we are no longer able to capture proximity effects.
Therefore, outputs tend to be blurry due to spatial averaging.

Algorithm \ref{alg:spatial} shows the implementation of the spatial channel.
Let $\tilde{T}$ be the occluded LST image as input, $f$ be the size of the local window, and $\theta^{*}$ be the maximum occlusion factor threshold for local filtering. 
We obtain interpolated image $\hat{T}_{sp}$ as follows:

\begin{algorithm}[H]
\caption{Computation of the spatial channel}\label{alg:spatial}
\begin{algorithmic}[1]
\Require $\tilde{T}, f, \theta^*$
\If{$\theta < \theta^*$} \Comment{local filtering}
\For{$\tilde{T}_{\bd{p}} \in \tilde{T}$}
\If{$O_{\bd{p}} = 1$}
\State $\hat{T}_{\bd{p}} \gets \frac{1}{\alpha} \sum_{\boldsymbol{q} \in N} G_{\sigma}(\lVert \boldsymbol{p} - \boldsymbol{q} \rVert) ~\bar{O}_{\boldsymbol{q}}~f_L(\boldsymbol{p}, \boldsymbol{q})~\tilde{T}_{\boldsymbol{q}} $\EndIf
\EndFor
\Else \Comment{global averaging}
\For{$\tilde{T}_{\bd{p}} \in \tilde{T}$}
\If{$O_{\bd{p}} = 1$}
\State $c \gets L_{\bd{p}}$
\State $\hat{T}_{\bd{p}} \gets \mu_c$  
\EndIf
\EndFor
\EndIf
\State \Return $\hat{T}^{sp} \gets \hat{T}$
\end{algorithmic}
\end{algorithm}

In our implementation, we set the values of the optimization parameters as $f = 75$ and $\theta^{*} = 0.5$. These values were selected based on extensive testing on a subset of images in our dataset to achieve the best performance.

\begin{figure}[H]
\centering
\includegraphics[width=0.9\linewidth]{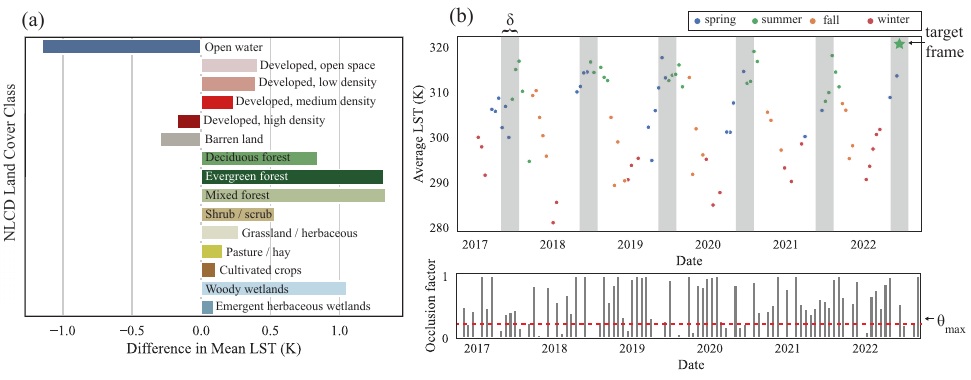}
\caption{\textbf{Temporal dynamics of LST per land cover class. }(a) Changes in average LST for each land cover class between Dec 5th and 21st, 2018, in Houston. 
The average LST for different classes drifts in different magnitudes and sometimes in different directions. 
(b) We select reference frames for the temporal channel via two conditions: (i) in the same temporal cycle as the target frame, indicated by gray stripes, and (ii) under a maximum tolerable occlusion factor, $\theta_{max}$.
$\delta$ denotes the temporal bracket duration. }
\label{fig:temporal_motivation}
\end{figure}

\subsection{Temporal Channel} \label{subsec:temporal}
In this section, we show how to generate an interpolated prediction using temporal information. 
We call this method the \textit{temporal channel}, which is complementary to the spatial channel defined in Sec.~\ref{subsec:spatial}. 
As seen in Fig.~\ref{fig:temporal_motivation}, the intuition behind the temporal channel is that objects in the same land cover class tend to exhibit similar thermal dynamics over time. 
The temporal channel involves four steps: (a) select a set of frames as a reference, (b) preprocess each reference frame via the spatial channel, (c) apply linear adjustments to each reference image, and (d) interpolate occluded regions based on the set of adjusted reference frames.

\textbf{Reference frame selection:}
We select reference frames based on two conditions: (i) seasonality and (ii) cloud occlusion.
The goal is to identify suitable reference images that can be used to accurately reconstruct occluded LST.

For seasonality, we take into account the temporal offset between the occluded target image $\tilde{T}$ and other available images $\tilde{T}'$.
Due to temporal continuity, the previous and the next LST sample over the same region are good reference points.
Constrained by the revisit cycle of Landsat 8, the previous and next samples are taken 16 days before and after the target date.
In addition to the immediate temporal neighbors, we also leverage the predictable seasonal variations in LST signals. This allows us to include the previous and the next sample acquired in other years as potential references. 
In our actual implementation, we increase the selection limit to samples collected 2 cycles prior to or after the target date in each year, defining the \textit{temporal bracket duration} $\delta$ as 2. 
See vertical gray stripes in Fig.~\ref{fig:temporal_motivation}(b) for a visual illustration.

The second condition of reference frame selection is the occlusion factor $\theta$. 
We have observed that minimally occluded reference frames lead to smaller interpolation errors.
Therefore we only consider selecting reference frames whose $\theta$ is below a certain maximum tolerable threshold, $\theta_{\text{max}}$ (see red dotted line in Fig.~\ref{fig:temporal_motivation}(b)).
We call the set of images satisfying these two conditions candidate reference frames $\mathcal{T}^\ast_{ref}$.
Within $\mathcal{T}^\ast_{ref}$, we prioritize selecting images that are captured closer to the target frame in time, as measured by the temporal difference $\Delta(| \tilde{T}' - \tilde{T}|)$. 
To achieve this, we choose a subset of $n$ frames from $\mathcal{T}^\ast_{ref}$ with the lowest $\Delta(| \tilde{T}' - \tilde{T}| )$.
We call this selected subset of reference frames $\mathcal{T}_{ref}$.

\textbf{Spatial channel pre-processing:}
Given a selection of reference frames $\mathcal{T}_{ref}$, we apply the computation of spatial channel from Fig.~\ref{subsec:spatial} to produce a set of spatially complete reference frames.
As a result, each pixel contains either observed or interpolated temperature data.
Note that the imposed constraint on $\theta_{max}$ means that the spatial channel only needs to interpolate a minimal amount of occlusion. 

\textbf{Linear adjustments:}
After pre-processing $\mathcal{T}_{ref}$, we apply a linear adjustment to all pixels in each class. Fig.~\ref{fig:temporal_motivation}(a) shows that the changes in the mean LST (denoted as $\mu_c$) can differ drastically between classes. 
Although the selection of reference frames helps mitigate these discrepancies, some adjustments are still necessary.
Specifically, for each reference frame, we add the difference in $\mu_c$  (i.e.,~$\Delta \mu_c$) between two dates to all pixels belonging to the corresponding land cover class. 
$\Delta \mu_c$ visually translates to the length of each bar in Fig.~\ref{fig:temporal_motivation}(a).
We repeat this process for all images in $\mathcal{T}_{ref}$.

\textbf{Reference frame-based interpolation:}
After the previous two steps, we obtain a set of linearly-adjusted reference frames. 
The interpolated LST image, $\hat{T}^{temp}$, is the average of each linearly adjusted reference frame.
This interpolation step combines the information from multiple reference frames to produce a spatially complete and temporally consistent estimate of the occluded LST image.

 Algorithm \ref{alg:temporal} shows the implementation of the procedures above. Let $n$ be the number of reference frames to be selected, $\delta$ be the temporal bracket duration, and $\theta_{max}$ be the maximum tolerable occlusion factor. We compute the interpolated image $\hat{T}_{temp}$ as follows:

\begin{algorithm}[H]
\caption{Computation of the temporal channel}\label{alg:temporal}
\begin{algorithmic}[1]
\Require $\tilde{T}, n, \delta, \theta_{max}$
\State $\mathcal{T}^\ast_{ref} = \{\tilde{T}' \in \mathcal{T} | \delta_{\tilde{T}'} < \delta \land \theta_{\tilde{T}'} < \theta_{max}\}$  \Comment{Reference set conditions}
\State $\mathcal{T}_{ref} = \displaystyle \min_{\Delta(|\tilde{T}' - \tilde{T}|)}(\tilde{T}' \in \mathcal{T}^\ast_{ref})_{(n)}$ \Comment{Reference frame selection}
\For{$\tilde{T}' \in \mathcal{T}_{ref}$}
\State $\hat{T}' \gets$ SP($\tilde{T}', f, \theta_{min}$)  \Comment{Preprocess via spatial channel}
\State $\Delta \mu_c = \frac{\sum_{\bd{p} \in P_c} |\tilde{T}_{\bd{p}} - \tilde{T}'_{\bd{p}}|}{|P_c|}~\forall c$
\State $\tilde{T}_{\bd{p}}' \gets \tilde{T}_{\bd{p}}' + \Delta \mu_c~\forall \bd{p} \in P_c~\forall c$ \Comment{Linear adjustments}
\EndFor
\State \Return $\hat{T}^{temp} \gets \text{average}(\mathcal{T}_{ref})$ \Comment{Interpolate}
\end{algorithmic}
\end{algorithm}

 We set $n = 3$, $\delta = 2$, and $\theta_{\text{max}} = 0.1$. SP($\cdot$) denotes the function for spatial channel defined in Algorithm \ref{alg:spatial}, and $f$ and $\theta_{\text{min}}$ at line 4 follow their default values from Sec.~\ref{subsec:spatial}. 
The subscript $\cdot_{(n)}$ used in line 2 signifies the first $n$ elements of the sequence, arranged according to the minimization criterion.
$P_c$ is the set of all pixel locations where the corresponding land cover class is $c$, and $| P_c| $ is the cardinality of set $P_c$.

\subsection{Estimate LST via Weighted Average}

Following the previous two subsections, we acquire initial predictions $\hat{T}^{sp}$ from the spatial channel and $\hat{T}^{temp}$ from the temporal channel. The final interpolated LST, denoted as $\hat{T}$, is calculated as the weighted average of the two initial predictions:
\begin{equation}
 \label{eq:weighted_prediction_bt}
 \hat{T} = w \cdot \hat{T}^{sp} + (1 - w) \cdot \hat{T}^{temp}
\end{equation}
 where we set the weight $w = 1 - \theta$. 
Note that for a minimally occluded image (a small $\theta$), more weight is assigned to the spatial channel. 
Contrarily, for a severely occluded image (a large $\theta$), fewer neighboring pixels are available for the spatial channel, and more weight is assigned to the temporal channel accordingly. 

\section{Results}
\label{sec:5_results}
\begin{figure}[ht]
\centering
\includegraphics[width=0.9\linewidth]{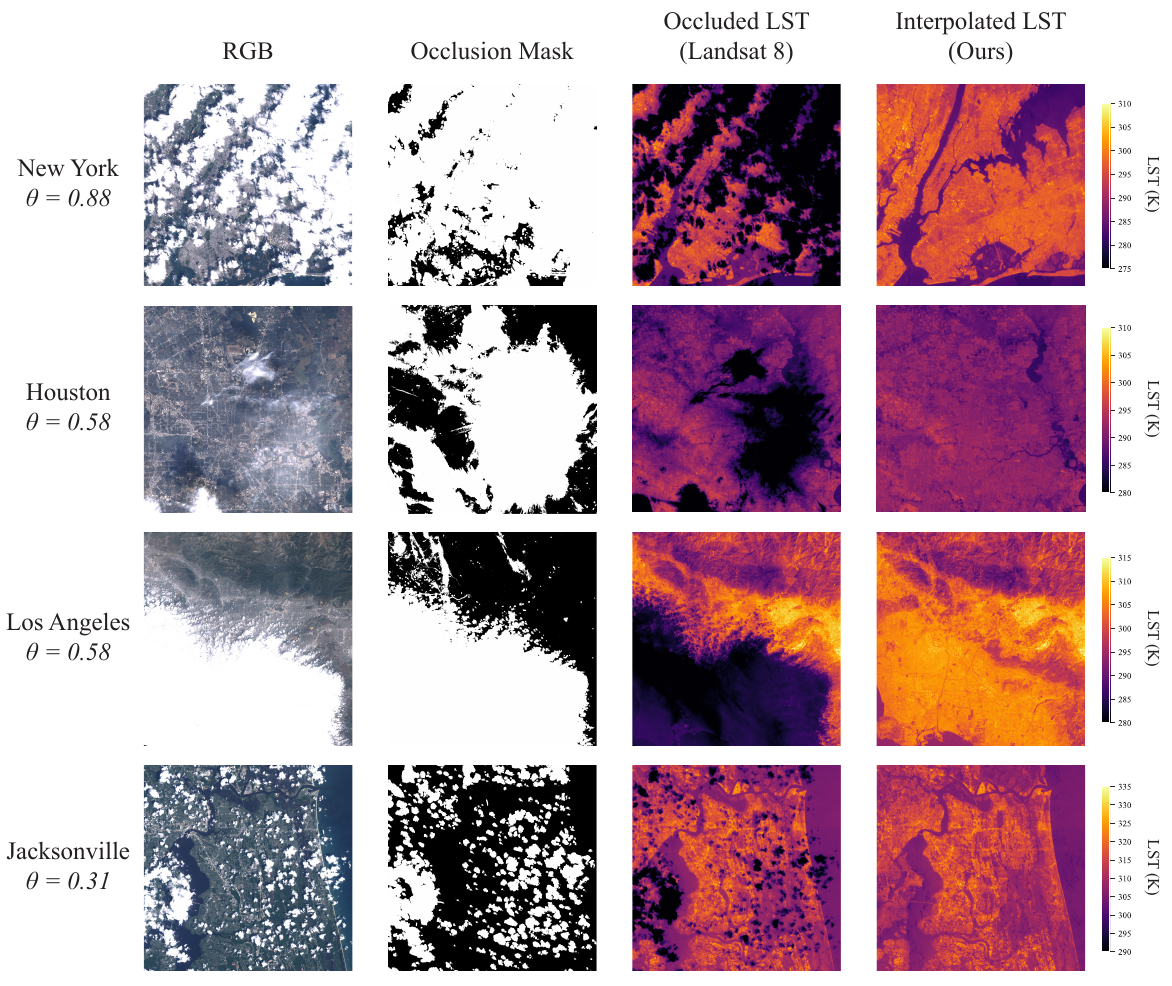}
\caption{\textbf{Results.} 
ISLAND performs robust interpolation under a variety of surface land cover and cloud occlusion conditions.
We show interpolated LST alongside LST inputs, occlusion masks, and RGB images for visual reference.
The fraction of pixels occluded is denoted as $\theta$.}
\label{fig:results}
\end{figure}

\subsection{Public Data Products} \label{subsec:public_data}

We deploy ISLAND across 20 regions in the United States. These regions are
selected for having the highest populations, as reported in the 2020 U.S.
Census~\citep{bureau_2023}.\footnote{For each region, we manually define a
polygon roughly covering the metro region for each city. Due to the 32 MB limit
per image download from Earth Engine, the polygon may not encompass the entire
metro region for some cities, such as New York.}
We collect Landsat data from 2019--2023 and NLCD 2021 release using the data compilation process described in Sec.~\ref{sec:3_data}.
We then use ISLAND to predict cloud-contaminated LST for each region. 
ISLAND produces interpolated LST at a spatial resolution of \SI{30}{\meter} every 16 days\footnote{Subject to data availability and requires $\theta < 0.99$.} for each observation region. 
Our public dataset is available on the NHERI DesignSafe
Cyberinfrastructure\footnote{https://doi.org/10.17603/ds2-3rf5-sd58}~\citep{rathje2017designsafe}.

Fig.~\ref{fig:results} shows a set of interpolated LST maps produced by ISLAND. We choose a diverse set of examples consisting of different land cover and cloud occlusion conditions. ISLAND reconstructs spatially complete LST maps under a variety of conditions, from lightly occluded by thin cirrus clouds (Houston) to heavily occluded by optically thick clouds (New York, 88\% occluded). Our model performs well across regions with different land cover characteristics, from dense urban settings (New York and Los Angeles) to diverse wetlands (Jacksonville). 

\begin{table}[!htbp]
    \begin{center} 
    \caption{\textbf{Simulation evaluation metrics.} 
    We simulate cloud contamination by artificially occluding LST images. 
    We report MAE and RMSE on artificially occluded areas for different cities.
    We also vary the size of each occlusion $s$, and the maximum number of occlusions $n$.
    ISLAND (marked as M1) is compared against other models in an ablation study. 
    Legends: 
    \textbf{M1}: ISLAND; \textbf{M2}: spatial channel only; \textbf{M3}: temporal channel only; \textbf{M4}: No NLCD input; \textbf{M5}: Naive average.
}
    \resizebox{0.85\columnwidth}{!}{\begin{tabular}{ccc|ccccc|ccccc}
        \multicolumn{3}{c}{Parameters} & \multicolumn{5}{c}{MAE (K) $\downarrow$} & 
        \multicolumn{5}{c}{RMSE (K) $\downarrow$} \\
        \midrule
        City & $s$ & $n$ & \textbf{M1} & M2 & M3 & M4 & M5 & \textbf{M1} & M2 & M3 & M4 & M5\\
        \midrule
        Houston & 250 & 10 & 1.88 & 2.19 & \textbf{1.82} & 2.49 & 4.26 & 2.43 & 2.85 & \textbf{2.33} & 3.10 & 5.05\\
        Houston & 750 & 3 & 2.00 & 2.39 & \textbf{1.87} & 2.57 & 3.97 & 2.53 & 3.05 & \textbf{2.38} & 3.25 & 4.71\\
        Jacksonville & 75 & 2 & \textbf{1.47} & 1.57 & 1.53 & 2.14 & 3.89 & \textbf{1.88} & 2.07 & 1.88 & 2.61 & 4.51\\
        Phoenix & 500 & 1 & 1.96 & 2.08 & \textbf{1.36} & 2.18 & 2.63 & 2.62 & 2.78 & \textbf{1.81} & 2.88 & 3.33\\
        New York & 100 & 2 & 1.25 & 1.37 & \textbf{1.14} & 2.11 & 3.82 & 1.65 & 1.84 & \textbf{1.42} & 2.63 & 4.28\\
\bottomrule
    \end{tabular}} \label{tab:sim_eval}
    \end{center}
\end{table}

\subsection{Simulation Evaluation}\label{subsec: sim_eval}

To evaluate the performance of our LST reconstruction method, we simulate cloud contamination by artificially occluding Landsat LST images. The added occlusions occupy a set of $s \times s$ rectangular regions. We set LST pixel values to zero and occlusion mask to True in these regions. To evaluate, we choose four urban regions with different climatological and surface land cover conditions, as seen in Table~\ref{tab:sim_eval}. For each region, we apply up to $n$ occlusion regions of size $s \times s$ pixels for all available LST images from 2019--2023. We use NLCD 2021 release as input for our simulation evaluation. For LST images with actual cloud occlusion (i.e.,~\emph{real} occlusion), we place artificial occlusions alongside real occlusions and ensure no overlap. The evaluation metrics are computed only in artificially occluded areas by comparing them to the original LST images. We report the mean absolute error (MAE) and root mean squared error (RMSE) in Table~\ref{tab:sim_eval}. M1 refers to our model ISLAND, while M2 - M4 refers to other models discussed in \ref{subsec: ablation}.

Table~\ref{tab:sim_eval} suggests that the RMSE typically ranges from 1.65--\SI{2.62}{K} for a variety of urban regions and occlusion scenarios. The first two rows of Table~\ref{tab:sim_eval} indicate that more occlusions generally lead to larger reconstruction errors. ISLAND performs well in dense urban regions (such as New York City) and diverse wetlands (such as Jacksonville). Our simulation evaluation indicates that ISLAND is robust in performance and is able to generalize reasonably well to different regions in the U.S. and under different occlusion characteristics.

\subsection{Ablation Study on Simulation Data} \label{subsec: ablation}

To further demonstrate the effectiveness of our model, we perform an ablation study~\footnote{Not to be confused with the glaciological definition of ablation, which refers to the process of removing snow, ice, or water from a glacier or a snow field. Here, we use the artificial intelligence definition of ablation study, where certain components of a model are removed in order to gain a better understanding of the model's behavior. }, where we remove key components of our model and observe the impact of each of these components on the overall performance. Table~\ref{tab:sim_eval} shows a list of models. M1 refers to our full model, ISLAND. In M2, we exclude the temporal channel and only keep the spatial channel. In M3, we discard the spatial channel and only keep the temporal channel. Note that M1 is a weighted average of M2 and M3, following Eq.~(\ref{eq:weighted_prediction_bt}). M4 is the same as ISLAND, except we remove NLCD land cover labels as input. M5 adopts a simplified approach where missing values are replaced with the average pixel value within an image without considering their spatial distribution or land cover class.

Table~\ref{tab:sim_eval} shows that the use of NLCD land cover labels (M1) leads to better reconstruction performance over the model without NLCD as input (M4). Such difference is more pronounced in dense urban regions, such as New York, and heterogeneous regions, such as Jacksonville.

When confronted with heavier occlusions (large $\theta$), M3 consistently emerges as the top-performing model. 
This finding highlights the effectiveness of relying solely on the temporal channel in such challenging scenarios. 
Conversely, for occlusions of low to moderate severity, the spatial channel exhibits stronger performance.
By appropriately favoring the temporal prediction in the presence of heavy occlusions, and the spatial prediction for lighter occlusions, our weighting scheme (Eq.~(\ref{eq:weighted_prediction_bt})) allows for adaptability to varying occlusion levels, enhancing the model's robustness and accuracy in capturing LST under diverse conditions.

\begin{figure}
\centering
\includegraphics[width=0.75\linewidth]{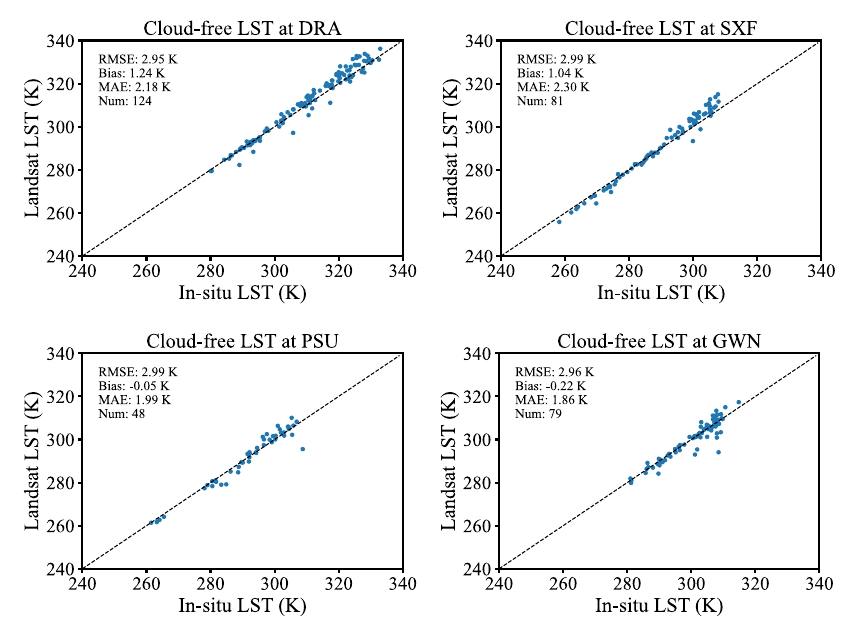}
\caption{\textbf{Underlying uncertainties in Landsat LST retrieval.}
Scatter plots of Landsat 8 LST versus in situ LST at four SURFRAD sites during 2013--2020 under \emph{clear-sky} conditions.}
\label{fig:surfrad_clear}
\end{figure}

\begin{figure}
\centering
\includegraphics[width=0.75\linewidth]{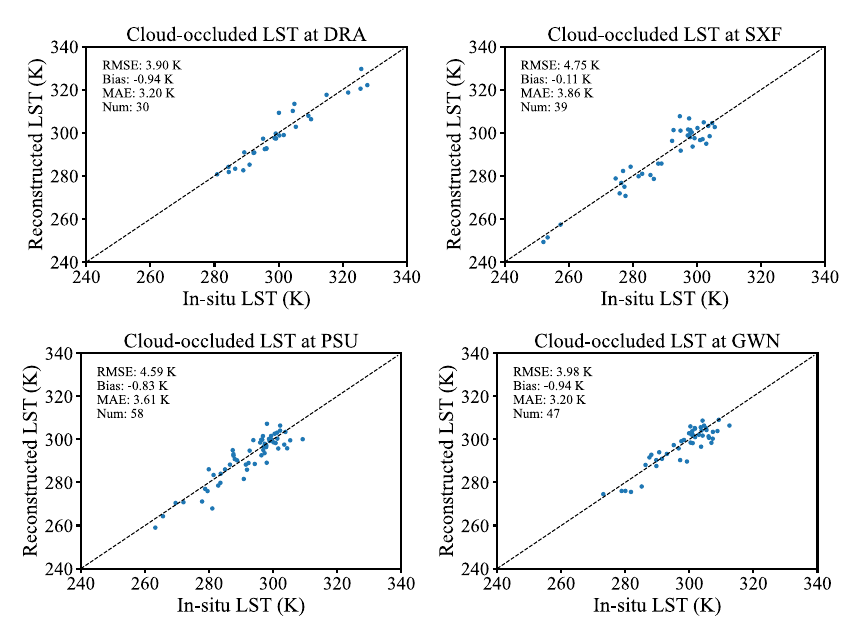}
\caption{\textbf{In situ validation.}
Scatter plots of ISLAND reconstructed LST versus in situ LST at four SURFRAD sites during 2013--2020 under \emph{cloudy-sky} conditions.}
\label{fig:surfrad_val}
\end{figure}

\subsection{In situ Evaluation} \label{subsec: in_situ_eval}

In this section, we evaluate ISLAND reconstruction results against in situ LST measurements collected in four SURFRAD stations. Our in situ evaluation uses Landsat and SURFRAD data collected in 2013--2020 and NLCD 2016 release as inputs to ISLAND. In \ref{subsec: sim_eval}, simulation evaluation examines ISLAND's ability to reconstruct a theoretical clear-sky LST under artificial occlusion. However, studies have shown that clouds have a cooling effect on the shaded region~\citep{weng2014modeling}, and simulation data is unable to evaluate if our reconstruction algorithm accounts for this effect. As such, we use in situ measurements to evaluate reconstruction performance under cloud-sky conditions.

There are two primary disadvantages to evaluating ISLAND against SURFRAD in situ data. 
First, remote retrieval of LST, even in clear-sky conditions, has uncertainties. 
Sources of error include uncertainty in emissivity estimation, atmospheric compensation, etc. 
Further, a Landsat 8 pixel (at \SI{30}{\meter} resolution) is larger than the
field-of-view (FoV) of the SURFRAD instrument~\citep{malakar2018operational}.
Differences in FoV compounded with spatial heterogeneity in temperature at
SURFRAD sites~\citep{malakar2018operational} cause Landsat LST to further
deviate from in situ LST. Therefore, we report Landsat LST versus in situ LST under clear-sky conditions (without involving ISLAND) in Fig.~\ref{fig:surfrad_clear}. RMSE ranges from $2.95 -$\SI{2.99}{\kelvin} across four sites.\footnote{We choose not to include two other SURFRAD sites, BND and FPK, because the Landsat clear-sky RMSE at these sites is too high in our calculation, at \SI{3.55}{\kelvin} and \SI{4.78}{\kelvin} respectively.} The error here represents the underlying uncertainties of comparing Landsat LST with in situ LST, which is external to our reconstruction algorithm. 

The second disadvantage of SURFRAD evaluation is that all SURFRAD sites are located in rural, homogeneous areas. Recall that the primary advantage of ISLAND is the use of NLCD data, which is applicable to urban areas but not to SURFRAD sites. Unfortunately, there are no publicly available in situ LST validation sites located in urban regions. Despite these limitations, we report in situ validation results to build an understanding of how ISLAND performs under real cloud occlusion. 

Fig.~\ref{fig:surfrad_val} shows ISLAND reconstructed LST versus in situ LST under cloudy-sky conditions. We define a given data point as cloudy-sky if and only if the corresponding Landsat pixel is flagged as cloud, cloud shadow, or cirrus, according to CFMask. Reconstruction RMSE ranges from $3.90-$\SI{4.75}{\kelvin}, across four sites. Compared to clear-sky RMSE in Fig.~\ref{fig:surfrad_clear}, ISLAND introduces an additional $0.95-$\SI{1.76}{\kelvin} RMSE error across four sites, with the average additional RMSE being \SI{1.33}{\kelvin}.

Fig.~\ref{fig:surfrad_val} also shows that there is no systematic overestimation (i.e.,~positive bias) across all four SURFRAD sites, suggesting that ISLAND effectively accounts for the local cooling effects caused by clouds. We believe that the modeling of the local cooling effect is primarily driven by the spatial channel defined in \ref{subsec:spatial}. As clouds move, they also cool the surrounding neighborhood. Under the assumption that surface objects have some degree of thermal inertia, we believe that the spatial channel utilizes surrounding cooler pixels to account for the local cooling effect caused by clouds, thereby accurately predicting LST under cloudy-sky conditions. While simulation evaluation (Table~\ref{tab:sim_eval}) shows that using the temporal channel alone (M3) leads to better results than M1, in situ evaluation, however, suggests that using both spatial and temporal channels (M1, Fig.~\ref{fig:surfrad_val}) leads to better results. When using temporal channel only (M3), the RMSE at DRA, SXF, PSU, and GWN are \SI{3.96}{K}, \SI{5.07}{K}, \SI{4.88}{K}, and \SI{3.86}{K}, respectively. 

Finally, Fig.~\ref{fig:surfrad_val} includes reconstruction error for all cloudy-sky conditions except for dates with more than 99\% of pixels occluded ($\theta > 0.99$), demonstrating ISLAND's robustness under a wide variety of cloud occlusion scenarios, including severely occluded LST images.

\subsection{Applications} \label{subsec:applications}

In this section, we show a set of applications demonstrating the impact of ISLAND on a variety of LST applications in urban environments.

\begin{figure}
\centering
\includegraphics[width=0.85\linewidth]{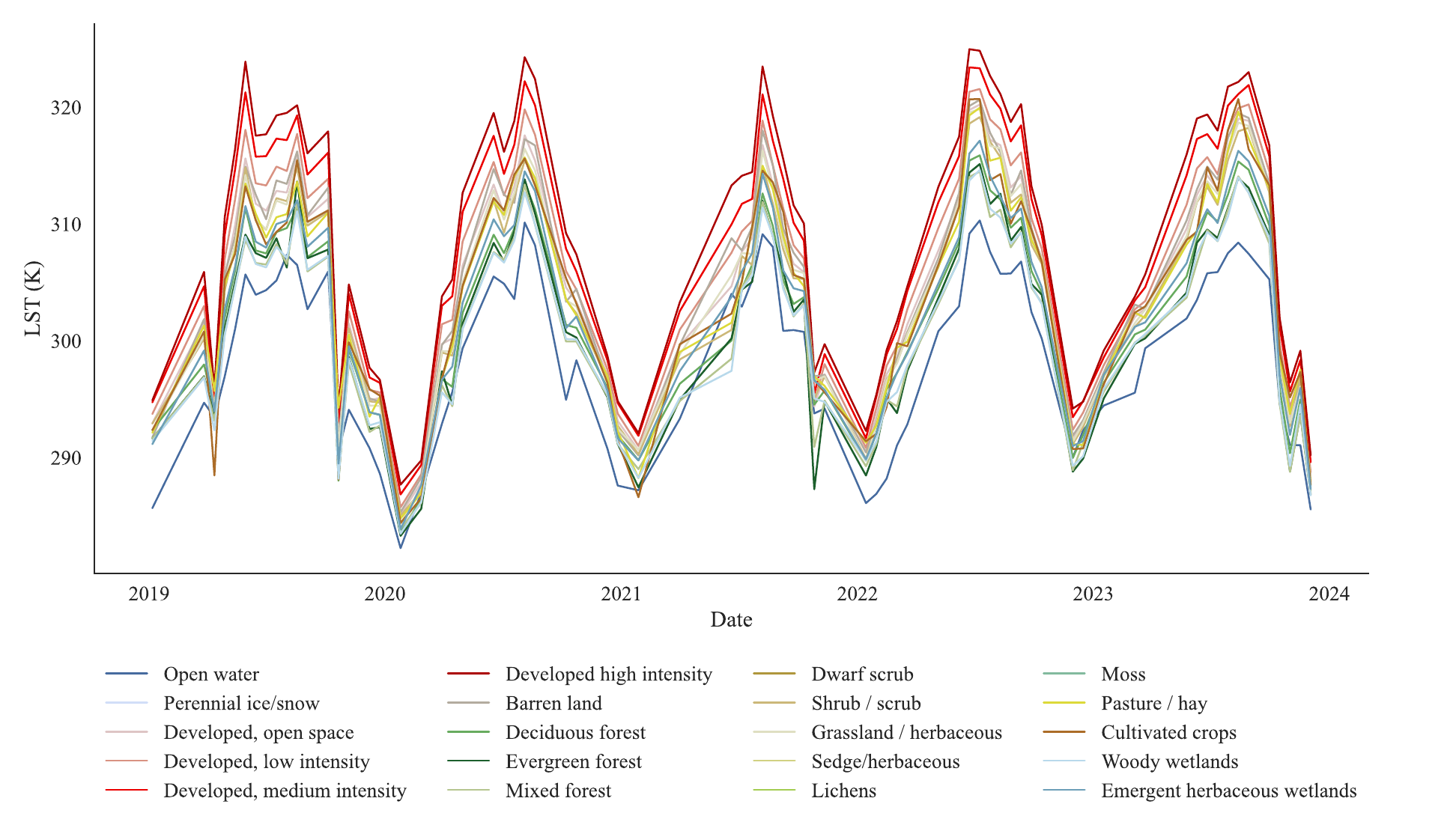}
\caption{\textbf{Progression of LST in Houston, 2019--2023.} This plot shows the mean LST for different NLCD land cover classes in Houston over the span of 5~years. By providing temporally consistent LST outputs, ISLAND enables time-series analysis of LST. }
\label{fig:temp_trends}
\end{figure}

\subsubsection{Deriving surface temperature trends}

The robustness demonstrated in Sec.~\ref{subsec: sim_eval} -- \ref{subsec: in_situ_eval} and the compelling results displayed in Fig.~\ref{fig:results} underscore the model's ability to generate accurate interpolated LST values across a wide range of conditions, with the only constraint being that the occlusion factor $\theta < .99$. As highlighted in Sec.~\ref{sec:1_intro}, previous studies investigating changes in land surface temperature through remote sensing have been constrained by limited observational conditions, restricting their analyses to a fraction of dates characterized by minimal cloud occlusion \citep{sobrino_evaluation_2013,baiocchi_remote_2017,huang_investigating_2019,gomez-martinez_multi-temporal_2021}. However, with the introduction of ISLAND, these limitations become obsolete, granting access to a significantly expanded set of operational LST data, particularly in urbanized regions.

Beyond the production of interpolated image outputs, ISLAND enables the examination of temporal variations in LST on daily-to-seasonal timescales. By reconstructing skillful LST maps for the majority ($\theta < .99$) of observation dates, ISLAND enables the comparison of thermal behaviors for a given region across time, at a relatively dense sampling rate of every 16 days, and encompassing diverse land cover types. To illustrate this capability, Fig.~\ref{fig:temp_trends} showcases the evolution of surface temperature in Houston. Each colored line represents the average LST for a particular land cover class for all grid cells over Houston. Pronounced temperature seasonality is evident in the time series from 2019--2023, with clear changes in seasonal structure from year to year. The ability to partition surface temperatures retrieved from different land cover types reveals differences of up to $15^\circ$C between forested, water-covered surfaces and urban developed surfaces (e.g.,~open-water vs. developed high intensity). With a 16-day temporal resolution, one can evaluate temperature distributions and variances over different land cover classes in different seasons. ISLAND facilitates an in-depth investigation of the temporal dynamics of surface temperature within any region located in the CONUS, providing valuable insights for climatological and ecological analyses.

\begin{figure}
\centering
\includegraphics[width=0.9\linewidth]{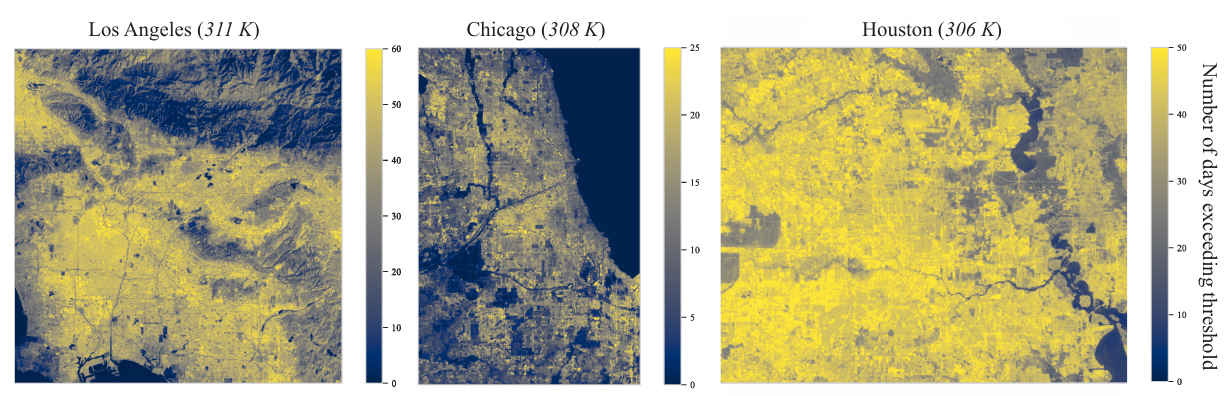}
\caption{{Urban heat island effects in Los Angeles, Chicago, and Houston.} Each figure panel shows the \textit{number of days} in which the LST at each pixel location exceeds a predetermined threshold (shown in parentheses). The temperature threshold for each city is chosen based on the definition of \textit{Extreme Danger} from the National Weather Service heat index.\label{fig:hotzones}}
\end{figure}

\subsubsection{Urban heat island effects}

Another key application of our model is the ability to study urban heat island
effects (UHIE)~\citep{moller_2022} at high spatiotemporal resolution.
UHIE refers to the phenomenon of urban areas being significantly warmer than their surrounding rural areas.
UHIE is primarily driven by the differences in thermal absorption between different materials.
For example, grass- or water-covered surfaces tend to have lower temperatures than concrete and asphalt.
By providing high spatial and temporal resolution LST outputs, ISLAND offers a novel data product for identifying, studying, and monitoring UHIE in major U.S. metropolitan areas.
 Fig.~\ref{fig:hotzones} shows maps of three of the largest U.S. metropolitan areas, Los Angeles, Chicago, and Houston, where the pixel values indicate the number of days surpassing a region-specific temperature threshold.
The thresholds are selected based on the definition of \textit{Extreme Danger}
from the National Weather Service (NWS) heat index~\citep{rothfusz1990heat}. 
The NWS heat index is a function of both temperature and relative humidity.
The Comparative Climatic Data (CCD-2018)~\citep{noaa2020comparative} provides
the morning annual average relative humidity ($R$) for each city. 
We select the LST threshold for each region as the minimum ambient dry bulb temperature that meets the NWS Extreme Danger criteria for the city's corresponding $R$.
The range of observations is 4.5~years, at a sampling rate of once per 16 days. 
Higher values in the frequency maps indicate a more frequent occurrence of UHIE. 
These frequency maps are available at \SI{30}{\meter} resolution and can be easily computed using our public dataset. 
From an urban planning perspective, these UHIE frequency maps offer a powerful tool for enhancing our understanding of how land cover choices influence micro-climates, heat extremes, and the associated health risks. 
By providing insights into the spatial distribution and frequency of UHIE, these maps can inform decision-making processes regarding urban development and land cover management, aiming to mitigate the adverse effects of heat on public health and well-being.

\subsubsection{Social Vulnerability \& Urban Heat Stress}

As illustrated in the last two applications, ISLAND facilitates the development of comprehensive datasets of LST and UHIE. The developed datasets will enable better characterization of heat exposure and its impacts on social, infrastructure, and environmental systems. A representative example application would be to investigate inequities in urban heat exposure. 
Given the health, well-being, and quality of life implications of urban heat,
and initiatives like Justice 40~\citep{whitehouseJustice40Initiative}, which
call for federal climate investments to be directed to environmental justice
communities, understanding the equities in urban heat exposure can centrally
guide prospective investments. 
For example, quantifying inequities in exposure to urban heat will help design adaptation measures such as increasing vegetation cover or guiding urban planning, among others.

\begin{figure}
\centering
\includegraphics[width=0.9\linewidth]{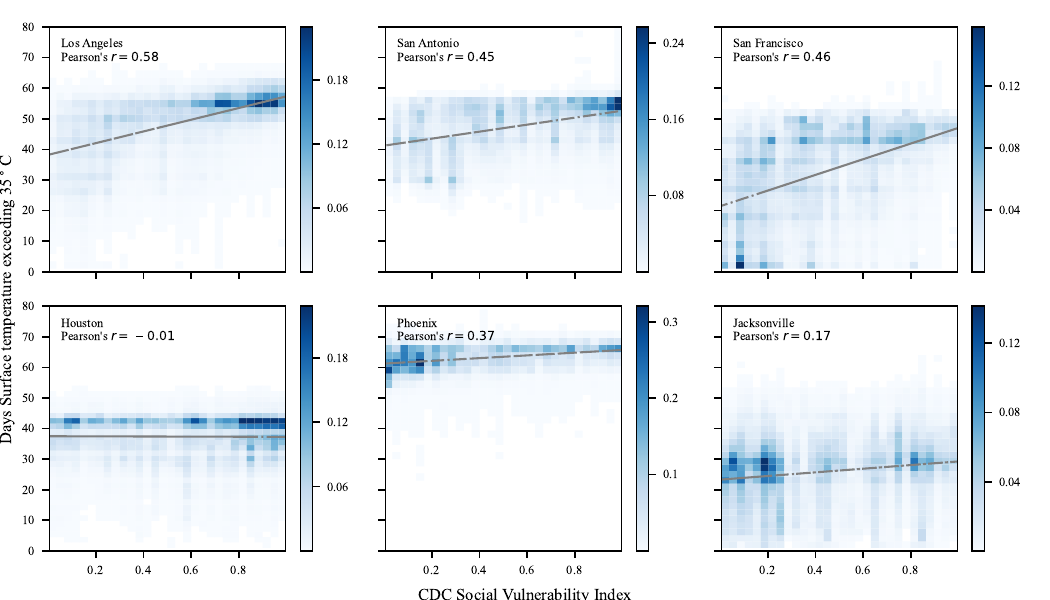}
\caption{\textbf{Equity in exposure to urban heat island effect.} Each figure panel shows a density plot mapping the relationship between the \textit{number of days} in which the LST of a pixel exceeded $35^\circ$C and their social vulnerability measured using CDC social vulnerability index. Here we restrict our analysis to residential areas within each city.}
\label{fig:svi}
\end{figure}

The distribution of UHIE for residential areas and the social vulnerability of
the exposed population for a few cities are shown in Fig.~\ref{fig:svi}. Here,
social vulnerability is measured using the Centers for Disease Control and
Prevention Social Vulnerability Index (CDC SVI)~\citep{cdc_cdc/atsdr_2020}. The
CDC SVI measures social vulnerability on a scale of 0 (least vulnerable) to 1
(most vulnerable), taking into account socioeconomic status (e.g.,~housing cost
burden), household characteristics (e.g.,~civilian with a disability), racial
and ethnic minority status (e.g.,~Hispanic, Alaska Native), housing type and
transportation factors (e.g.,~no vehicle). The latest available residential
land use data (2016) from~\citet{mc2022gridded} are used to identify
residential regions. Additionally, UHIE is calculated as the number of days
that a pixel (resolution of \SI{30}{\meter} $\times$ \SI{30}{\meter})
exceeds a land surface temperature threshold of $35^\circ$C~(308.5
K). Only pixels with at least one day of temperature over the threshold are
considered for the analysis.

From Fig.~\ref{fig:svi}, many cities show a systemic inequity in heat exposure. For example, in Los Angeles, socially vulnerable communities are exposed to high urban temperatures compared to less socially vulnerable communities (Pearson's correlation, $r=0.58$). A similar trend can be seen in cities such as San Antonio ($r=0.45$) and San Francisco ($r=0.46$), San Jose ($r=0.44$), and New York ($r=0.34$). In contrast, cities such as Houston ($r=-0.01$), Dallas ($r=-0.10$), Jacksonville ($r=0.17$), and Fort Worth ($r=0.08$) show no or negligible inequity in urban heat exposure. Factors ranging from vegetative cover to colocation with industrial or commercial locations might influence urban heat exposure. By providing reliable and more complete datasets to quantify UHIE, this study will allow for a better understanding of the factors influencing heat exposure and, as a result, will aid in developing strategies to mitigate urban heat stress.

While this section only provides three case studies, the proposed method facilitates many applications requiring high-resolution site-specific data. Some example applications include designing building envelopes and Heating, Ventilation, and Air Conditioning (HVAC) systems,
investigating the influence of thermal stresses on infrastructure aging and deterioration, power system reliability, and energy demand shifts,
or even high-resolution weather and natural hazard modeling that accounts for fine-scale surface temperature effects.  

\section{Discussion \& Conclusions}
\label{sec:6_discussion}
\subsection{Assumptions and Limitations} \label{subsec:limitations}

Despite the demonstrated validity and effectiveness of ISLAND, the model does contain important assumptions and limitations.

\textbf{Limitations of NLCD land cover labels:} 
As stated in Fig.~\ref{fig:overview_flowchart}, our model uses the NLCD land cover data.
Specifically, we use the NLCD 2021 release~\citep{dewitz_national_2021} for our
public data products, which most accurately reflect the state of land cover
labels in the year 2021.
Our Landsat LST inputs, in contrast, span from 2019--2023.
Within this observational period, NLCD 2021 alone does not reflect changes in land cover due to urban expansion, meandering, or coastal erosion.
NLCD releases every three years, and there are versions available for 2019, 2016, etc.
Although a fraction of observation dates could have potentially benefited from using the 2019 release instead of 2021, we chose not to implement computation using multiple NLCD versions for simplicity.
Despite using only the NLCD release for 2021, we still observe a significant advantage in terms of reconstructed LST when employing the NLCD dataset as input, as seen in 
Table~\ref{tab:sim_eval}.

\textbf{Reliance of accurate cloud masks}:
Another assumption of our model is that all clouds are correctly labeled. Our algorithm relies on accurate cloud labeling, and errors in cloud labeling would most likely propagate to LST estimations. 
Recall from Fig.~\ref{fig:overview_flowchart} that we utilize
CFMask~\citep{zhu2012object} to determine if a given pixel is affected by
cloud, cloud shadow, or cirrus.
CFMask algorithm can produce inaccurate cloud labels, leading to erroneous LST reconstructed values.
Cirrus is a category of clouds known to be challenging to
detect~\citep{Qiu_Zhu_Woodcock_2020}.
Row two of Fig.~\ref{fig:results} shows that unidentified cirrus in the lower left corner leads to visible discontinuities in the reconstructed LST image.
Unidentified cirrus pixels seem to affect the spatial channel more than the temporal channel.
Moreover, it is difficult to reliably detect clouds in snow-covered terrain due
to the spectral similarity between cloud and
snow~\citep{Stillinger_Roberts_Collar_Dozier_2019}.
Consequentially, in a simulated evaluation, we observe significantly higher RMSE in Denver, a region with prolonged snow coverage.
Finally, in some cases, we also observe that optically bright and thermally cold buildings are falsely labeled as clouds, though this mislabeling is rare. 
The increasing prevalence of white roofing applied to increase urban albedo and
decrease the UHIE may worsen the future uses of ISLAND~\citep{fayad_2021}.
It is important to address these challenges and improve cloud labeling techniques to ensure the accuracy of our model's predictions.

\textbf{Errors external to interpolation}:
As mentioned in Sec.~\ref{subsec: in_situ_eval}, remote LST retrieval itself has underlying uncertainties, even in clear-sky conditions.
Other sources of error include, but are not limited to, sensor calibration,
atmospheric profiles, and emissivity
estimation~\citep{li_satellite-derived_2013}.
Profiling and mitigating these types of errors are areas of active
research~\citep{li_satellite-derived_2013} in the field of remote sensing; such
improvements are beyond the scope of this paper.

\subsection{Transferability and Scalability}

We designed ISLAND to be easily accessible to the broader research community. 
All required inputs of our model listed in Fig.~\ref{fig:overview_flowchart} are
acquired from publicly available sources and are extracted from Google Earth
Engine~\citep{gorelick2017google} and its Python API, geemap~\citep{wu_2020}.
Since our model is essentially based on a set of filters, we do not require massive computing power or GPU acceleration. 
For context, it takes roughly 2~min to process one image 
on a 12-core CPU (AMD Ryzen 5900) with 32~GB of DRAM.

Currently, our model is only available to the CONUS region, constrained by the NLCD dataset. 
In Fig.~\ref{subsec:future_improvement}, we discuss potential avenues for expanding the operational region.

\subsection{Potential Improvements and Future Opportunities for ISLAND} \label{subsec:future_improvement}

\textbf{Extending area of study:}
In this paper, visual results (Fig.~\ref{fig:results}), simulation evaluations (Table~\ref{tab:sim_eval}), and demonstrated applications are all focused on urban regions.
While in situ evaluation (Fig.~\ref{fig:surfrad_clear} and \ref{fig:surfrad_val}) are conducted on SURFRAD sites, which offers an indicator of model performance on regions with relatively high land cover homogeneity (see Fig.~\ref{appendix_nlcd} for visual illustrations), extensive future testing is required to better quantify performance outside of urban settings.

In Sec.~\ref{subsec:limitations}, we showed that the use of the NLCD dataset currently restricts our analysis to the CONUS region.
In theory, we can potentially expand to other regions where there are appropriate land cover labels.
For example, the Copernicus CORINE Land Cover
dataset~\citep{Buchhorn_Lesiv_Tsendbazar_Herold_Bertels_Smets_2020} would
facilitate the extension of the model over Europe; the China land cover dataset
(CLCD)~\citep{Yang_Huang_2021} is also available, enabling research over the
Asian continent. It is important to note that the performance of ISLAND is
impacted by spatiotemporal resolution, diversity, and accuracy of land cover
labels.
Therefore, careful consideration should be given to the suitability and quality of the land cover datasets when expanding the study area beyond CONUS.
Additional testing is required to quantify ISLAND performance across different datasets and cities outside of CONUS with different land cover characteristics.

\textbf{Improving temporal resolution:}
Higher temporal resolution for LST products is instrumental for downstream operational studies. 
Our model uses Landsat 8~\citep{landsat8} data to achieve one LST
reconstruction every 16 days.
Recently, the Landsat program launched a companion satellite, Landsat
9~\citep{landsat9}, carrying a nearly identical TIRS as Landsat 8. 
Landsat 8 and Landsat 9 are phased eight days apart. 
By incorporating data from both satellites, we can reduce the time gap between consecutive satellite visits to just eight days. 
This enables us to capture LST measurements at a higher temporal resolution.

Another avenue to increasing the temporal resolution is to use satellite
products with shorter revisit cycles. For example, the
Sentinel-3~\citep{donlon2012global} program provides a temporal resolution of
at least once per day (at the equator). Unfortunately, the spatial resolution
of their LST products is lower than that of Landsat 8 (e.g.,~Sentinel-3 at
\SI{1}{km}). 
Additional benchmarking on the selected data source is required before applying ISLAND, as performance could vary based on spatio-temporal resolution. 

\textbf{Incorporating deep learning:}
The basis of our model is a set of filters designed around adjacency and temporal properties of thermal signatures. 
As seen in Algorithms \ref{alg:spatial} and \ref{alg:temporal}, these filters are \textit{hand crafted} to explicitly represent these relationships. 
While we clearly demonstrated the effectiveness of ISLAND through qualitative and quantitative analysis, we acknowledge that a well-designed deep learning algorithm has the potential to achieve even better performance. 
Here, we highlight a few examples.
Firstly, deep learning models have the capability to capture complex inter-class relationships between different land cover labels. 
This could enhance the overall accuracy of our interpolator. 
Secondly, a dynamic spatial channel that adapts based on occlusion characteristics could be incorporated, allowing the model to better handle varying cloud cover conditions. 
Additionally, an optimized weighting scheme, an improved cloud detection filter, and an updated NLCD land cover dataset to account for changes in land cover could be integrated into a deep learning framework.
Lastly, integrating additional data sources, such as other satellite data and
ground-based observations, could further reduce reconstruction errors, and
there are existing deep learning techniques~\citep{han2024time} to perform data
fusion.
Given the complexity of the problem and the non-linear nature of
LST~\citep{wu_spatially_2021}, deep learning is a suitable direction for future
work, but designing and training a deep learning framework might require
extensive research.

\subsection{Advancing Existing State-of-the-Art LST Estimates}

\begin{figure}
\includegraphics[width=0.95\linewidth]{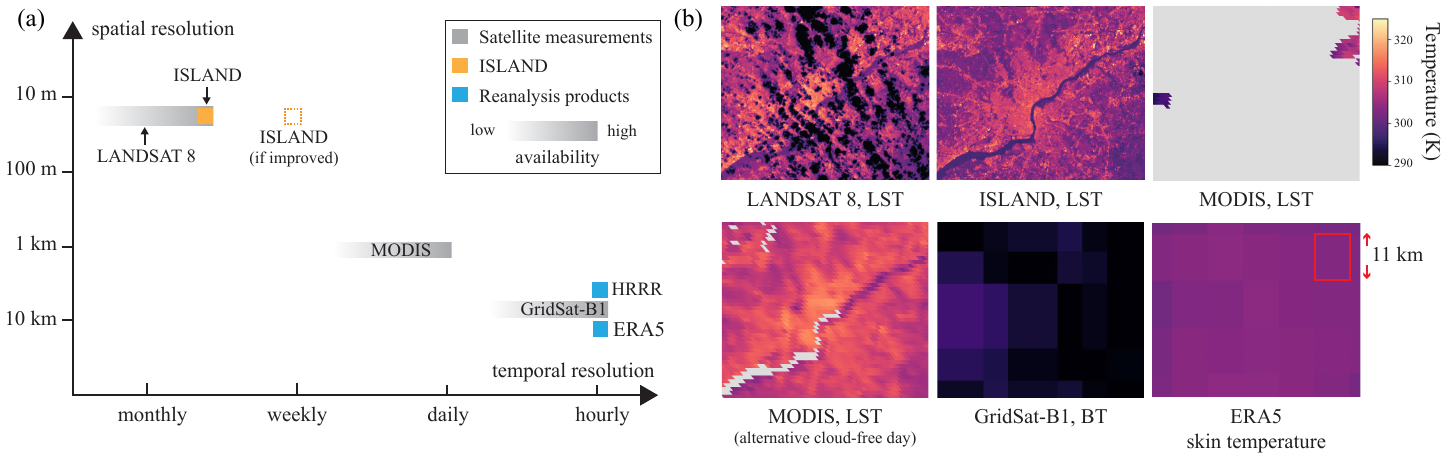}
\caption{\textbf{Contribution in the context of spatial--temporal resolutions.}
(a) ISLAND is derived from Landsat 8, which collects LST with 16-day revisit cycles, but the actual usable temporal resolution is much lower due to cloud contamination.
ISLAND improves the usable temporal resolution of Landsat 8 LST data to approach its 16-day limit.
If Landsat 9 data is to be integrated into a future variant of ISLAND, in theory, it could provide nearly weekly coverage (marked as if improved).
Other methods provide higher temporal resolution but with substantially lower spatial resolution.
(b) Six different LST products for Philadelphia, which highlight the spatial resolutions and the impact of cloud contamination across different methods.}
\label{fig:contribution}
\end{figure}

In this paper, we showed that a large fraction of Landsat measurements are occluded by clouds. As a consequence, the actual usable temporal resolution of Landsat is significantly reduced, falling below once per month when subject to frequent cloud occlusions. The role of ISLAND is to mitigate cloud contamination in LST images, maximizing its usable temporal resolution.

Indeed, the addition of ISLAND represents an advance over existing LST products, as shown in Fig.~\ref{fig:contribution}.
Generally speaking, there is a trade-off between spatial resolution and temporal resolution for satellite LST products.  
Amongst all available satellite measurements, Landsat 8 (along with the later-launched Landsat 9) offers the best spatial resolution at \SI{30}{\meter}, with 16-day revisit cycles.
Operating in sun-synchronous orbits (SSO), Landsat 8 has the advantage of providing global coverage and maintaining time-constant illumination conditions of the observed surfaces (except for seasonal variations).
The MODIS program~\citep{Wan_2013}, consisting of a pair of satellites named
Aqua and Terra, offers a much higher resolution at daily revisit cycles but
provides LST data at a much lower spatial resolution of \SI{1}{\kilo\meter}.
Satellites in geostationary orbits do provide higher temporal resolution at the expense of spatial resolution and global coverage.
For example,   GridSat-B1~\citep{Knapp_2011,Knapp_2014} provides brightness
temperature (BT) data at a resolution of \SI{7792}{\meter}.

In addition to satellite-based methods, climate reanalysis data provides an alternative approach to obtaining LST.
These products are generally designed to maintain the best possible physical
and temporal consistency and require prohibitive computational
costs~\citep{hakim_2016}.
The spatial resolution of these products is not comparable to satellite-based
methods; HRRR~\citep{james2022high} provides climate data at \SI{3}{\km} and
ERA5~\citep{ECMWF_2019} at around \SI{11}{\km}.
 Fig.~\ref{fig:contribution}(b) provides a visual comparison of the spatial resolution of LST from ERA5 and ISLAND. 
Due to relatively low spatial resolution, the urban spatial structure over Philadelphia is indistinguishable in ERA5 skin temperature fields. In contrast, our method effectively removes cloud contamination and produces a high-resolution reconstruction of LST at \SI{30}{\meter} resolution.

As shown in Fig.~\ref{subsec:applications}, many downstream applications generally benefit from increased spatial and temporal resolution.
For dense urban settings, high spatial resolution is particularly desirable, making the Landsat data a preferred choice.
In Fig.~\ref{subsec:future_improvement}, we discussed the potential of incorporating measurements from Landsat 9 to further enhance the temporal resolution to 8 days. 
This advancement would bring us closer to achieving consistent weekly measurements at a spatial resolution of \SI{30}{\meter}.
Such a combination of high spatial and temporal resolution of LST data is instrumental to our understanding of urban areas.

\subsection{Conclusions}

This paper introduces ISLAND, a novel model designed to address the issue of cloud occlusion in satellite LST images. ISLAND removes occlusion by estimating LST pixel values through a set of spatio-temporal filters. These filters account for the land cover class, resulting in higher LST reconstruction accuracy. ISLAND addresses a fundamental limitation of LST retrieval via remote sensing, thereby dramatically increasing the number of serviceable LST images via a robust mechanism to mitigate cloud contamination. These improvements enable nearly bi-weekly coverage of LST at \SI{30}{\meter} resolution over the CONUS region, a large advance over previously available LST products derived from remote sensing. We show ISLAND can operate in a variety of land cover types and cloud occlusion scenarios in both simulations and in situ evaluations. Overall, ISLAND provides a promising framework for a multitude of scientific applications that require high-resolution, frequent observations of LST, including but not limited to (1) urban heat island effects, (2) derivation of surface temperature trends, and (3) social vulnerability and urban heat stress.

\section*{Acknowledgement}

The authors gratefully acknowledge the support of this research by the National Science Foundation (NSF) award numbers 1652633 and 2107313. The contributions of Pranavesh Panakkal and Jamie E. Padgett were partially supported by NSF award number 2227467. Any opinions, findings, conclusions, or recommendations expressed in this paper are those of the authors and do not necessarily reflect the views of the sponsors.  

\appendix

\section{Additional Details on NLCD Land Cover}
\label{appendix_nlcd}
\begin{figure}
\centering
\includegraphics[width=0.7\linewidth]{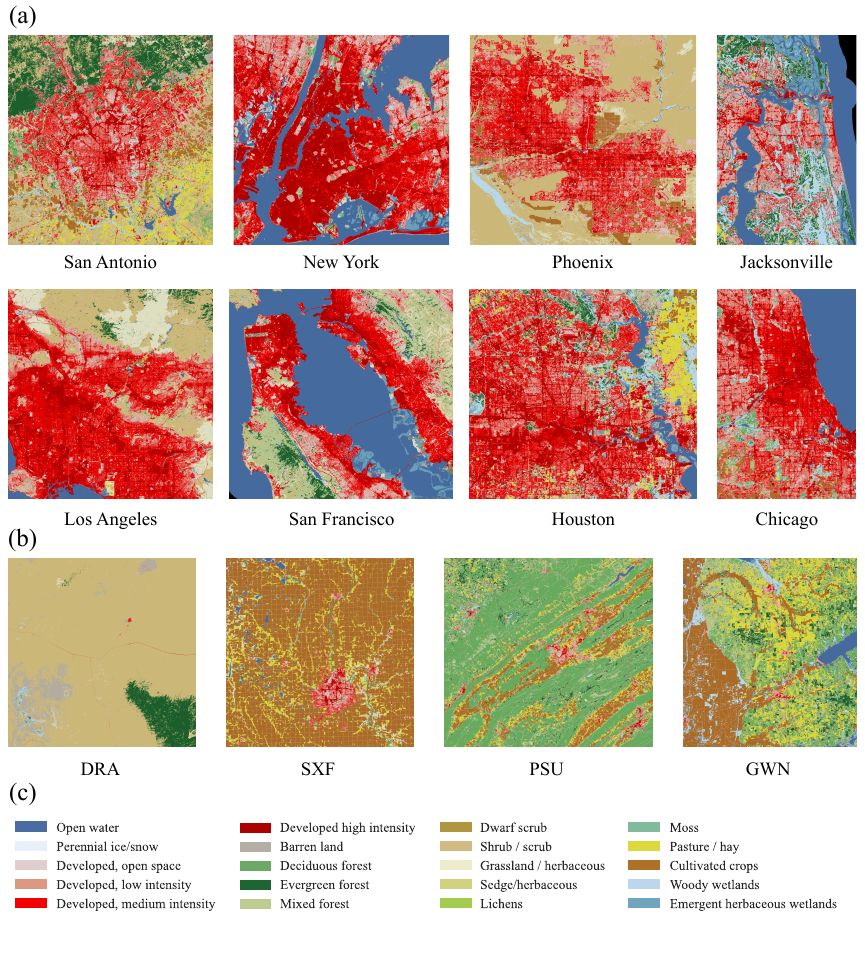}
\caption{\textbf{Visualization of NLCD land cover maps.} (a) NLCD 2021 land cover maps four US cities referenced in this paper. (b) NLCD 2016 land cover maps around SURFRAD sites referenced in Fig.~\ref{fig:surfrad_clear} and \ref{fig:surfrad_val}. (c) NLCD land cover legends.}
\label{fig:nlcd_maps}
\end{figure}

Fig.~\ref{fig:nlcd_maps} shows the visualization of the NLCD maps referenced in this paper. Fig.~\ref{fig:nlcd_maps}(a) shows urban NLCD maps for cities in our visual results (Fig.~\ref{subsec:public_data}), simulation evaluation (Sections~\ref{subsec: sim_eval}--\ref{subsec: ablation}), and illustrated applications (Fig.~\ref{subsec:applications}), while Fig.~\ref{fig:nlcd_maps}(b) shows the NLCD maps around four SURFRAD sites for our in situ evaluation (Fig.~\ref{subsec: in_situ_eval}). Refer to Fig.~\ref{fig:nlcd_maps}(c) for the legend of NLCD classes. 

The heterogeneity of land cover types in urban settings is clearly reflected in Fig.~\ref{fig:nlcd_maps}(a). As discussed in Fig.~\ref{fig:spatial_motivation} and \ref{fig:temporal_motivation}, LST is closely related to land cover; as such, complex terrains in cities make LST reconstruction challenging. The difference in land cover distributions between urban regions is also evident in Fig.~\ref{fig:nlcd_maps}(a). New York City is characterized by high-density developments with very sparse natural landscapes. Houston, TX, has extensive urban and suburban developments with mixed land use, accompanied by agricultural land, forest, and water bodies on the outskirts. Jacksonville, FL, has large expanses of coastal wetlands. Phoenix, AZ, is an urban region surrounded by deserts and mountains with sparse vegetation. We chose the four regions to conduct our simulation evaluation (Table~\ref{tab:sim_eval}) to represent ISLAND's performance under a variety of settings. 

Our in situ validation targets are SURFRAD stations located in rural regions. Fig.~\ref{fig:nlcd_maps}(b) shows the relative homogeneous land cover types surrounding the SURFRAD stations, which is in stark contrast to the urban regions in Fig.~\ref{fig:nlcd_maps}(a).

\FloatBarrier

\bibliographystyle{apalike}
\bibliography{main}
\end{document}